\documentstyle[twoside,fleqn,npb,epsfig,axodraw]{article}
\newcommand{\AmS}{{\protect\the\textfont2
  A\kern-.1667em\lower.5ex\hbox{M}\kern-.125emS}}

\title{Deep inelastic lepton-hadron scattering as a test of
       perturbative QCD}

\author{W.L. van Neerven \address{Instituut-Lorentz, University of Leiden,
        P.O. Box 9506, 2300 RA Leiden, The Netherlands}}

\begin{document}

\begin{abstract}
Exploration of the small $x$ kinematic region by the HERA experiments 
led to a revival of some models which existed before the advent of 
Quantum Chromo Dynamics (QCD) as the theory of the strong interactions.
Predictions of these models for the deep 
inelastic structure functions are compared with those given by QCD.
Future experiments will concentrate on the large $x$-region and we will
discuss some issues which are important for the test of QCD. In particular
we emphasize the next-to-next-to-leading (NNLO) order analysis of the
structure functions and the determination of the strong coupling constant
$\alpha_s$. We also make some critical remarks about the relevance
of so called large corrections in the small and large $x$-region.
\end{abstract}

\maketitle

\section{Introduction}

\begin{figure}
\begin{center}
  \begin{picture}(175,130)(0,0)
    \ArrowLine(0,20)(70,30)
    \Line(90,30)(148,35)
    \Line(80,25)(160,25)
    \Line(90,20)(148,15)
    \Line(130,50)(160,25)
    \Line(130,0)(160,25)
    \GCirc(80,30){20}{0.3}
    \Photon(50,100)(72,48){3}{7}
    \ArrowLine(0,100)(50,100)
    \ArrowLine(50,100)(100,120)
    \Text(0,110)[t]{$l_1$}
    \Text(100,130)[t]{$l_2$}
    \Text(0,15)[t]{$H$}
    \Text(50,75)[t]{$V$}
    \Text(175,30)[t]{$`X'$}
    \Text(30,20)[t]{$p$}
    \Text(25,115)[t]{$k_1$}
    \Text(75,125)[t]{$k_2$}
    \Text(75,75)[t]{$\downarrow q$}
  \end{picture}
  \caption[]{Kinematics of deep inelastic lepton-hadron scattering.} 
  \label{fig:1}
\end{center}
\end{figure}
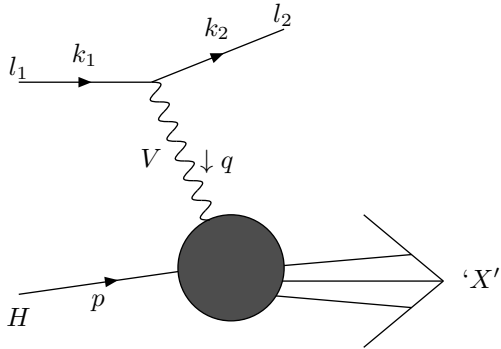

Deep inelastic lepton-hadron scattering is given by the following process
(see Fig. \ref{fig:1})
\begin{eqnarray}
\label{eqn:1}
l_1(k_1) + H(p) \rightarrow l_2(k_2) + `X'\,.
\end{eqnarray}
Here $`X'$ denotes any inclusive final hadronic state. The in and outgoing
leptons are represented by $l_1$ and $l_2$ respectively and the hadron is 
denoted by $H$.
On the Born level the reaction proceeds via the exchange of one of
the vector bosons $V$ of the standard model which are given by 
$\gamma$, $Z$ and $W^{\pm}$. The kinematic variables needed in this paper
are defined by
\begin{eqnarray}
\label{eqn:2}
q^2=-Q^2 < 0 \,,\quad \nu = \frac{p.q}{M}\,, \quad x=\frac{Q^2}{2M\nu}\,,
\nonumber\\[1ex]
W^2=(p+q)^2=2M\nu-Q^2+M^2\,,
\end{eqnarray}
where $W$ represents the CM energy of the virtual boson-hadron system.
The most interesting quantities under study are the structure functions 
denoted by $F_i(x,Q^2)$ ($i=2,3,L$) which provides us with information 
about the strong interaction between the quarks and gluons.
Besides these structure functions one can also study
other observables for instance those which are related to the production of 
jets. However due to a lack of space  we will limit ourselves
to a discussion of the structure functions only.

\section{QCD and alternative models}
Immediately after the first experiments carried out at SLAC \cite{bloom}
\footnote{Similar experiments were also carried out at DESY see \cite{bartel}} 
several models were proposed to explain the behaviour of the structure
functions. The most prominent among them are 
\begin{itemize}
\item[a.]
Light Cone Expansions \cite{brpr}
\item[b.]
Parton Model \cite{feyn}
\item[c.]
Vector Meson Dominance (VMD) \cite{saku}
\item[d.]
Regge Pole Model \cite{har}
\item[e.]
Dual Resonance Models
\end{itemize}
One could also add the BFKL \cite{bfkl} approach which appeared later 
on the scene than
the models above. It can be considered as a merger of ideas put forward
by the models mentioned under b and d.
\subsection{Light Cone Expansions}
\vspace*{3mm}
Suppressing Lorentz indices the structure function can be written as 
\begin{eqnarray}
\label{eqn:3}
&& F(x,Q^2)= 
\nonumber\\[1ex]
&& \frac{1}{4\pi M} \int d^4z \, e^{iq.z} \langle p \mid 
[J(z),J(0)] \mid p \rangle\,,
\end{eqnarray}
In the commutator above $J(z)$ represents the electro-weak current.
In the rest frame of the proton, with $p=(M,0)$, the exponent becomes
$q.z \sim \nu (z_0-z_3)-xMz_3$ for $\nu \gg M$. Using Fourier analysis 
the integral only gets contributions in the region $q.z = {\rm const.}$. 
Therefore the integrand is dominated by the space-time region \cite{brpr}
\begin{eqnarray}
\label{eqn:4}
\mid z_0-z_3 \mid < \frac{1}{\nu}\,, \quad z_0< \frac{1}{x~M} \,, \quad 
z_3 < \frac{1}{x~M} \,.
\end{eqnarray}
Furthermore because of Einstein microcausality the commutator of
the electro-weak currents vanishes for $z^2<0$. From the latter and
Eq. (\ref{eqn:4}) one infers that the
integrand is dominated by the light cone region $z^2< 1/Q^2$ provided
$Q^2$ is chosen to be sufficiently large. Therefore this variable is
the relevant scale for the Operator Product Expansion (OPE) and the
Parton Model which are the ingredients of perturbative QCD.
However there are other arguments in the literature \cite{iof}
leading to the claim that the distance between absorption and emission
of the virtual vector boson, given by $z_0 \sim z_3 = 1/xM$, is the relevant
scale. If $1/xM> 2R$, where $R$ denotes the radius of the hadron, the
virtual vector boson fluctuates into a hadronic system which interacts with 
the hadron $H$ in Fig. \ref{fig:1} so that one deals with hadron-hadron
collisions rather than vector boson-hadron scattering. 
When $x$ is sufficiently small the inequality above is satisfied
\footnote{ For the proton, where $R=5~{\rm GeV}^{-1}$, this would mean that
$x<0.1$}
and one enters the so called Regge region where $2M \nu \gg Q^2$. Note that
at this moment this inequality can be experimentally only 
realized when the vector boson is represented by the photon.
The inequality $1/xM> 2R$ is the justification of small $x$ 
physics which is at the basis of e.g. the Regge Pole Model and Vector Meson 
Dominance.

When the light cone region dominates one can make the Operator Product 
Expansion (OPE) \cite{brpr}
\begin{eqnarray}
\label{eqn:5}
&& [J(z),J(0)] 
 {\raisebox{-2 mm}{$\,\stackrel{=}{{\scriptstyle z^2 \sim 0 }}\, $} }
\nonumber\\[1ex]
&&\sum_{\tau} \sum_{N} C^{N,\tau}(z^2 \mu^2) O^{N,\tau}(\mu^2,0)\,,
\end{eqnarray}
where $\tau$ and $N$ denote the twist and spin of the operator $O^{N,\tau}$
respectively. Both the coefficient function $ C^{N,\tau}$ and the operator
$O^{N,\tau}$ are understood to be renormalized and $\mu$ denotes the
re-normalization scale which later on can be identified with the factorization
scale. The operator product expansion has been proven for 
some quantum field theories in the context of perturbation theory. Assuming
that it also holds in QCD one can express the $N$th moment of the structure
function as follows
\begin{eqnarray}
\label{eqn:6}
&& \int_0^1dx\,x^{N-1}\,F(x,Q^2)=
\nonumber\\[1ex]
&& \sum_{\tau} \left (\frac{M^2}{Q^2} 
\right )^{\frac{\tau}{2}-1} A^{(N),\tau}(\mu^2){\cal C}^{(N),\tau}
\left(\frac{Q^2}{\mu^2}\right) \,.
\end{eqnarray}
In momentum space the operator matrix element and the coefficient function are 
defined by
\begin{eqnarray}
\label{eqn:7}
A^{(N),\tau}(\mu^2)=\langle p \mid O^{N,\tau}(\mu^2,0) \mid p \rangle \,,
\end{eqnarray}
and
\begin{eqnarray}
\label{eqn:8}
{\cal C}^{(N),\tau}\left(\frac{Q^2}{\mu^2}\right)=\int d^4 z \,e^{iq.z} 
C^{N,\tau}(z^2 \mu^2) \,,
\end{eqnarray}
respectively.

\subsection{Parton Model}
\vspace*{3mm}
For leading twist, i.e. $\tau=2$, the results of the operator product 
expansion are reproduced by the Parton Model. The Bjorken scaling variable
$x$, defined in Eq. (\ref{eqn:2}), has a nice interpretation if one
describes lepton-proton scattering in an Infinite Momentum
Frame (IMF) \cite{feyn}. The latter is e.g. given by the following choice 
for the proton and vector boson momenta
\begin{eqnarray}
\label{eqn:9}
p=(P + \frac{M^2}{2P}, \vec 0_{\perp}, P) \,, \quad 
q=(\frac{M \nu}{P}, \vec q_{\perp}, 0) \,.
\end{eqnarray}
Writing the proton state $\mid p \rangle$ as a sum of multi parton states
i.e. $\mid p \rangle=\sum_{i=1}^n \mid k_i \rangle$ the momenta of the
latter can be expressed as
\begin{eqnarray}
\label{eqn:10}
k_i&=&(x_i P + \frac{k_{i\perp}^2+m_i^2}{2x_iP}, \vec k_{i\perp}, x_iP) \,,
\nonumber\\
\mbox{with} && \sum_{i=1}^n x_i =1 \,.
\end{eqnarray}
Using the IMF representation,
the life time of the multi parton state becomes
\begin{eqnarray}
\label{eqn:11}
\tau_{part}\sim \frac{1}{\sum_{i=1}^n k_{i0} - p_0}=
\nonumber\\[1ex]
\frac{2P}{\sum_{i=1}^n \frac{k_{i\perp}^2+m_i^2}{x_i}-M^2}\,.
\end{eqnarray}
If the interaction time of the vector boson with the partons is given by
$\tau_{int}\sim 1/q_0 = P/M \nu$ then the scattering becomes incoherent
for $\tau_{part} \gg \tau_{int}$ or
\begin{figure}
\begin{center}
  \begin{picture}(100,100)(0,0)
  \ArrowLine(45,60)(80,60)
  \ArrowLine(80,60)(100,0)

  \ArrowLine(47,53)(100,53)
  \ArrowLine(49,46)(100,46)

  \DashLine(80,45)(80,21){4}

  \ArrowLine(45,20)(100,20)
  \Photon(80,100)(80,60){2}{5}
  \CArc(0,40)(50,322,38)
  \CArc(80,40)(50,142,218)
  
  \Line(0,42)(31,42)
  \Line(0,38)(31,38)
  \Line(13,45)(16,40)
  \Line(13,35)(16,40)
  \Text(16,35)[t]{$p$}
  \Text(88,100)[t]{$q$}
  \Text(60,73)[t]{$k_1$}
  \Text(100,65)[t]{$k_2$}
  \Text(100,44)[t]{$k_3$}
  \Text(73,15)[t]{$k_n$}
\end{picture}
\caption[]{\sf Parton Model for deep inelastic lepton-hadron scattering 
          in an Infinite Momentum Frame.}
\label{fig2}
\end{center}
\end{figure}
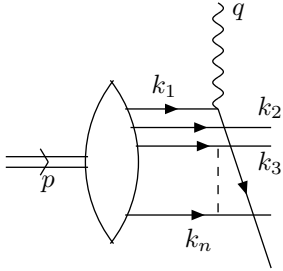

\begin{eqnarray}
\label{eqn:12}
2M\nu+M^2 \gg \sum_{i=1}^n \frac{k_{i\perp}^2+m_i^2}{x_i} \,.
\end{eqnarray}
The inequality above only holds if $x_i \not =0$ or $x_i \not =1$. From
these considerations one can derive the master formula for the structure
function
\begin{eqnarray}
\label{eqn:13}
F(x,Q^2)=\sum_{a=1}^m e_a^2 f_a(x) \,,
\end{eqnarray}
where $x$ is interpreted as the fraction of the proton's momentum carried
away by the parton struck by the virtual vector boson $V$ in Fig. \ref{fig:1}.
Furthermore $e_a$ and $f_a$ denote the parton charge and the parton density 
respectively where $a$ runs over the number
of active flavours in the initial state. Notice that in QCD the relation in 
Eq. (\ref{eqn:13}) will be modified after one has included radiative 
corrections.
From the discussion above it is clear that at small $x$ deep inelastic 
scattering becomes coherent. However the Parton Model does not give any
information at which value of $x$ this will happen. For this we have
to know much more about QCD. Notice that the Infinite Momentum
Frame picture is not the justification of the Parton Model. The latter can
be also formulated in a Lorentz covariant way.

The real justification for the light cone expansion and the Parton Model 
comes from QCD. This theory gives much more firm predictions when compared
with the models mentioned before and discussed hereafter. Let us
enumerate some of them
\begin{itemize}
\item[a.] The $Q^2$ dependence of $F(x,Q^2)$.
\item[b.] Sum rules of the type $\int_0^1 dx F(x,Q^2)$ which only depend
on the flavour group and the value of the strong coupling constant $\alpha_s$.
\item[c.]
Relations like the Callan-Gross relation \cite{cagr} 
which states that the longitudinal
structure function $F_L \sim 0 + O(\alpha_s)$.
\item[d.] Factorization theorems leading to the universality of parton
densities like $f_a$ in Eq. (\ref{eqn:13}). This enables us to predict 
distributions and cross sections of other hard processes than deep inelastic 
lepton-hadron scattering.
\item[e.] Rates and distributions of jets.
\end{itemize}
Here we want to emphasize that our theoretical understanding of
integrated quantities, like cross sections or the sum rules mentioned
under b, is much better than what is known about differential distributions
or even about the structure functions. This is all related to our ignorance
about the non-perturbative properties of QCD, which means that we
have to rely on models or even worse on parametrizations to describe the 
$x$-dependence of the parton
densities in Eq. (\ref{eqn:13}) or the fragmentation properties of
quarks and gluons into hadron jets. In particular the global analysis
of the parton densities requires many arbitrary free parameters. In
spite of the fact that these densities allow us to describe many hard processes
their arbitrariness impairs the predictive power of perturbative QCD.
It is for this reason that this theory has so many competitors in
particular in the small $x$ region where diffraction plays a prominent
role. Therefore better measurements of integrated quantities like
the sum rules above are indispensable to show the validity of QCD although we
realize that experimentally this is very hard to achieve. 

\subsection{Vector Meson Dominance}
\begin{figure}
\begin{center}
  \begin{picture}(100,100)(0,0)
  \Photon(0,70)(50,70){2}{5}
  \Line(50,72)(100,72)
  \Line(50,68)(100,68)
  \Line(50,72)(50,68)
  \Line(65,25)(85,25)

  \Line(67,27)(65,25)
  \Line(67,23)(65,25)

  \Line(83,27)(85,25)
  \Line(83,23)(85,25)

  \GCirc(75,10){10}{0.3}
  \Text(0,65)[t]{$q$}
  \Text(100,63)[t]{$p'$}
  \Text(90,10)[t]{$p$}
  \Text(75,40)[t]{$2~R$}
  \Text(100,83)[t]{$V$}
\end{picture}
\caption[]{\sf Photon $q$ fluctuating into a vector meson $V$ with
  momentum $p'$ and mass $m$.\\
$p=(M,0)$, $q=(\nu,0,0,\sqrt{\nu^2+Q^2})$,\\ 
$p'=(\sqrt{\nu^2+Q^2+m^2},0,0,\sqrt{\nu^2+Q^2})$.}
\label{fig3}
\end{center}
\end{figure}
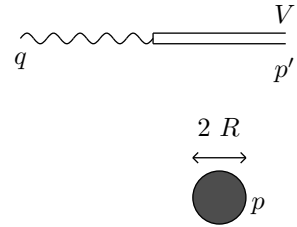

\vspace*{3mm}
In the case that the process in Fig. \ref{fig:1} is dominated by the photon
one can decompose it into a point-like and a hadron-like state
i.e. $\mid \gamma \rangle=\mid \gamma_{point} \rangle+\mid hadron \rangle$.
According to the parton model the hadronic state vanishes like $1/Q^2$.
However the argument for Vector Meson Dominance (VMD) is akin to the one
given for small $x$-physics presented below Eq. (\ref{eqn:4}). 
If $M\nu \gg Q^2 + m^2$ where $m$ is the mass of
the hadronic state the life time of the latter in the proton's rest frame
(see Fig. \ref{fig3}) is given by \cite{nieh}
\begin{eqnarray}
\label{eqn:14}
\tau_{had}\sim \frac{1}{p'_0-q_0} = \frac{2 \nu}{Q^2 + m^2} > 2\,R \,.
\end{eqnarray}
Choosing for the hadronic state the $\rho$-meson (lightest vector meson)
one can derive, using the proton radius $R=5~{\rm GeV}^{-1}$, the inequality
$\nu > 3 + 5Q^2$ which can be considered as the characteristic value of
$\nu$ for the diffractive region. The inequality above corresponds to $x < 0.1$
which belongs to the kinematic regime explored by HERA but not quite by
the old SLAC experiments. Before we discuss the predictions of VMD let us
first express the structure functions into the transverse $\sigma_T(Q^2,\nu)$
and longitudinal $\sigma_L(Q^2,\nu)$ photo absorption cross sections.
\begin{eqnarray}
\label{eqn:15}
F_2(x,Q^2)&=&\frac{K}{4 \pi^2 \alpha} \frac{\nu Q^2}{Q^2+\nu^2} 
(\sigma_T+\sigma_L) \,,
\nonumber\\[1ex]
F_L(x,Q^2)&=&\frac{K}{4 \pi^2 \alpha} \frac{Q^2}{\nu} \sigma_L \,,
\nonumber\\[1ex]
K&=&\nu - \frac{Q^2}{2M} \,.
\end{eqnarray}
In the literature there are two versions of VMD. In the older version
the hadronic
state was only represented by the $\rho$-meson \cite{saku}. Shortly after 
single vector meson dominance a second version was proposed called 
generalized VMD \cite{sasc}. In the latter case one has in 
principle an infinite number of
vector meson states which are produced in the process 
$e^++e^- \rightarrow `hadrons'$. The most general properties of VMD are
\begin{eqnarray}
\label{eqn:16}
\sigma_T(Q^2,\nu)&=& G(Q^2,\nu) \sigma_{\gamma p} (\nu) \,,
\nonumber\\[1ex]
\sigma_L(Q^2,\nu)&=& R(Q^2,\nu) \sigma_T(Q^2,\nu) \,.
\end{eqnarray}
In the formulae above $\sigma_{\gamma p}$ denotes the real photon-proton 
($q^2=0$) cross
section and $G$ is a model dependent function which can be inferred from
the process $e^++e^- \rightarrow `hadrons'$. 
The ratio $R$ in the two models
is given by
\begin{eqnarray}
\label{eqn:17}
R &{\raisebox{-2 mm}{$\,\stackrel{=}{{\scriptstyle Q^2 \gg m^2 }}\, $} }&
 \xi(\nu) \frac{Q^2}{m_{\rho}^2} \quad \mbox{VMD}\,, 
\nonumber\\[1ex]
R &{\raisebox{-2 mm}{$\,\stackrel{=}{{\scriptstyle Q^2 \gg m^2 }}\, $} }&
 \xi(\nu) \ln \left (\frac{Q^2}{m_0^2} \right ) \quad \mbox{gen. VMD} \,,
\end{eqnarray}
where $m_0$ is the averaged vector meson mass.
Further $\xi(\nu)$ is defined by
\begin{eqnarray}
\label{eqn:18}
\xi(\nu)=\frac{\sigma_{tot}(V~p)_{\lambda_V=0}}
{\sigma_{tot}(V~p)_{\lambda_V=\pm 1}} \,,
\end{eqnarray}
where $\sigma_{tot}(V~p)$ denotes the total cross section for the reaction
$V + p \rightarrow `X'$ and $\lambda_V$ represents the helicity of the vector
meson $V$. The main consequence of VMD is that at large $Q^2$ we obtain
$R \gg 1$ and $ F_L \sim F_2$. This is in contrast to
the prediction of the quark parton model which yields $R \ll 1$
and $ F_L \ll F_2$. The old SLAC data ruled out VMD with $V=\rho$ because
in this case $\xi$ is large (i.e. $\xi \sim 1.2$) and $R$ in Eq. (\ref{eqn:17})
rises linearly in $Q^2$. However generalized VMD works much 
better since the increase of $R$ is logarithmic and the fitted value for $\xi$ 
is much less (i.e. $\xi=0.171$) than the one found for $\rho$-meson dominance.
We have checked that the parametrisation in \cite{scsp} fits the data obtained 
for $F_L$ by the H1-group \cite{adloff} rather well in the region 
$10^{- 4}< x <5.10^{-4}$.
However at larger $x$ values i.e. for $x>0.01$ the model in \cite{scsp}
breaks down because both the transverse and longitudinal cross sections become
negative.
\subsection{Regge Pole Model}
\vspace*{3mm}
In the Regge Pole model the structure functions are parametrised as
follows
\begin{figure}
\begin{center}
  \begin{picture}(100,100)(0,0)
  \Photon(0,100)(48,70){2}{5}
  \Photon(100,100)(52,70){2}{5}
  \Line(0,0)(50,25)
  \Line(100,0)(50,25)
  \Line(0,5)(48,29)
  \Line(100,5)(52,29)
  \Line(48,70)(48,29)
  \Line(52,70)(52,29)
  \Line(48,70)(52,70)
  \Line(48,29)(52,29)

  \Line(48,65)(52,65)
  \Line(48,35)(52,35)

  \Line(48,60)(52,60)
  \Line(48,40)(52,40)

  \Line(48,55)(52,55)
  \Line(48,45)(52,45)
  \Line(48,50)(52,50)

  \Line(17,18)(25,15)
  \Line(22,8)(25,15)

  \Line(83,18)(75,15)
  \Line(78,8)(75,15)

  \Text(90,50)[t]{$i=P,A_2,f,f'$}
  \Text(0,95)[t]{$q$}
  \Text(100,95)[t]{$q$}
  \Text(25,8)[t]{$p$}
  \Text(75,8)[t]{$p$}
\end{picture}
\caption[]{\sf Regge pole exchange in forward Compton scattering.}
\label{fig4}
\end{center}
\end{figure}
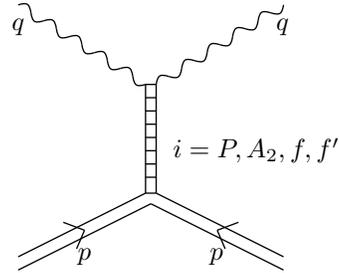

\begin{eqnarray}
\label{eqn:19}
F_k(x,Q^2)=\sum_i \, \beta_{k,i}(Q^2) \left ( \frac{\nu}{\nu_o} \right )^
{\alpha_i(0)-1} \,.
\end{eqnarray}
Here $\alpha_i(0)$ represents the Regge trajectory at zero momentum transfer
since we deal with forward Compton scattering $\gamma^* + p \rightarrow 
\gamma^* + p$. Further $i$ runs over all trajectories of the particles
which can be exchanged between the photon and the proton among which the most 
prominent are the pomeron $P$ and the mesons $A_2,f,f'$ (see Fig. \ref{fig4}). 
Since the real photon-proton cross section behaves like
$\sigma_{\gamma p}(\nu)\sim (\nu/\nu_0)^{\alpha_i(0)-1}$ a comparison between 
Eq. (\ref{eqn:19}) and
Eqs. (\ref{eqn:15}), (\ref{eqn:16}) reveals that VMD is a special case
of the Regge Pole Model providing us with an explicit expression for
$\beta_{k,i}(Q^2)$. However the Regge Pole Model does not predict scaling
unless one makes the additional assumption \cite{har}
\begin{eqnarray}
\label{eqn:20}
\beta_{k,i}(Q^2)=b_{k,i}\left ( \frac{Q_0^2}{Q^2} \right )^{\alpha_i(0)-1}\,,
\end{eqnarray}
so that one obtains
\begin{eqnarray}
\label{eqn:21}
F_k(x)\sim \sum_i \, b_{k,i} x^{1-\alpha_i(0)} \,.
\end{eqnarray}
In \cite{cudo} one has obtained a good fit to the data for $F_2$ in the range
$x<0.07$ and $0<Q^2<2000~{\rm GeV/c}^2$ provided one allows for
a $Q^2$ dependence of $b_{2,i}$. Moreover in addition to the
soft pomeron $P$ ($\alpha_P(0)=1.08$) one also has to add a hard pomeron 
$P'$ contribution
($\alpha_{P'}(0)\sim 1.4$). The assumption in Eq. (\ref{eqn:20}) has
a curious implication. This is revealed if we take the $N$th moment
of Eq. (\ref{eqn:21}) which is equal to
\begin{eqnarray}
\label{eqn:22}
\int_0^1 dx\,x^{N-1} F_k(x,Q^2)=\sum_i \frac{ b_{k,i}(Q^2)}{N-\alpha_i(0)}\,.
\end{eqnarray}
We will now show that this prediction is at variance with perturbative QCD. 
In the latter case the $N$th moment is given by
\begin{eqnarray}
\label{eqn:23}
\int_0^1 dx\,x^{N-1} F_k(x,Q^2)=
\nonumber\\[1ex]
\sum_{a=q,g} f_a^{(N)}(\mu^2) {\cal C}^{(N)}_{k,a}\left 
(\frac{Q^2}{\mu^2} \right ) \,.
\end{eqnarray}
Here $\mu$ represents the factorization as well as the renormalization scale. 
Since in the
Regge Pole Model $\alpha_i(0)$ is independent of $Q^2$ the location of
the singularities in the $N$-plane (see Eq. (\ref{eqn:22})) does not depend on
$Q^2$. In the scaling parton model where ${\cal C}^{(N)}=1$ the
Regge pole behaviour is also exhibited by the parton densities $f_a^{(N)}$.
However in perturbative QCD new poles will appear in the $N$-plane 
because of the behaviour 
\begin{eqnarray}
\label{eqn:24}
{\cal C}^{(N)}_{k,a} \sim \left (\frac{\alpha_s(\mu^2)}{\pi} \right )^m
\gamma_{ab}^{(N)} \ln \frac{Q^2}{\mu^2}\,.
\end{eqnarray}
The anomalous dimensions $\gamma_{ab}^{(N)}$ contain poles in the $N$-plane
different from those given by the Regge Pole Model. For example we have
\begin{eqnarray}
\label{eqn:25}
\gamma_{ab}^{(N)} \sim \frac{1}{N-1},\,\frac{1}{N},\,\frac{1}{N+1}
\qquad \mbox{etc.}
\end{eqnarray}
From Eq. (\ref{eqn:24}) we infer that the appearance of the pole terms
on the right hand side of Eq. (\ref{eqn:23}) depends on the value of $Q^2$
which is in contradiction with the assumption in the Regge Pole Model.
Notice that the singularities at $N=1$ disappear if the pole terms of
the type $1/(N-1)^l$ are re-summed in all orders of perturbation theory
using the BFKL equation \cite{bfkl}. This follows from the BFKL characteristic 
function 
\begin{eqnarray}
\label{eqn:26}
\chi \Big (a_s\gamma_{gg}(N) \Big)=\frac{N-1}{6 a_s} \,,
\qquad a_s=\frac{\alpha_s}{\pi}\,,
\end{eqnarray}
which does not contain a singularity in $\gamma_{gg}(N)$ at $N=1$. 
Unfortunately such 
an equation is not known for the other poles given by $N \ge 0$. 
However in principle there is no contradiction between perturbative QCD
and the Regge Pole Model provided one drops the assumption made
in Eq. (\ref{eqn:20}) leading to Eq. (\ref{eqn:21}). Therefore the Regge
poles show up in the angular momentum plane but not in the $N$-plane 
(moment-plane). Notice that $N$ is the quantum number associated with
the angular momentum in the four dimensional Euclidean plane. Hence
we conclude that the assumption made in Eq. (\ref{eqn:20}) (see \cite{har})
only works in the case of the old scaling parton model but not for QCD.
Hence we disagree with the conclusions of \cite{cudo} that Regge poles
contradict the singularity structure in the $N$-plane predicted by perturbative
QCD. Moreover we want to emphasize that the singularities at $N=1$
in $\gamma_{ab}^{(N)}$ (\ref{eqn:25}), which are called the leading poles,
pose no harm for the convergence of the perturbation series. First, as has
been shown in a model \cite{blne1}, the residues of the sub-leading 
singularities at $N=0,-1,-2 \cdots$ are much larger that the one corresponding
to the leading pole. Therefore the latter does not dominate the 
asymptotic behaviour of the coefficient function for $z \rightarrow 0$.
The second reason can be traced back to the structure function, appearing
in Eq. (\ref{eqn:23}), which can be written as a convolution of parton 
densities and coefficient functions in Bjorken $x$-space. Since these 
densities $f_a(x/z)$ vanish in the region $z \sim x$ the effect of the
leading poles in the coefficient function, which behave in $z$-space like 
$(\ln^m z)/z$, will be considerably
reduced. A systematic study regarding the phenomenological meaning of
these so called leading poles or small $x$-terms is presented in \cite{blvo}.

\section{Issues in QCD which are relevant for large $x$ and $Q^2$}
\vspace*{3mm}

\subsection{NNLO analysis of the structure functions}
\vspace*{3mm}
If one would like to have a test of perturbative QCD which can be performed
on the same level of accuracy as achieved in $e^+~e^-$ physics it is necessary
to get a next-to-next-to-leading order (NNLO) description of the
structure functions. Notice that some quantities like $R(e^+~e^- \rightarrow
`X')$ or $\Gamma(Z \rightarrow hadrons)$ are already known up to order 
$\alpha_s^3$ so that they are even one order higher in accuracy than the NNLO
expression of $F_k(x,Q^2)$. The latter can be used to check the QCD prediction
for its  evolution with respect to $Q^2$ and to extract the 
value of $\alpha_s$ which can be compared with the results obtained from
the LEP and SLC experiments. The ingredients for the NNLO analysis are
the order $\alpha_s^2$ corrected coefficients \cite{zn} and the order
$\alpha_s^3$ corrected anomalous dimensions (splitting functions). However
the three-loop contributions to the anomalous dimensions $\gamma_{ab}^{(N)}$
are only known for $N \le 10$ \cite{lvr} except for the leading 
$n_f$ part, where
$n_f$ denotes the number of flavours, which is computed in \cite{begr}.
For the computation of the integrals corresponding to the three-loop
graphs one has to make a thorough study of the transcendental functions
which show up in the splitting functions $P_{ab}(z)$ and the anomalous 
dimensions $\gamma_{ab}^{(N)}$.
Typical functions which appear in $P_{ab}(z)$ are the Nielsen integrals 
corresponding to finite and infinite harmonic sums in  $\gamma_{ab}^{(N)}$.
A study of Nielsen integrals and harmonic sums
is made in \cite{blku} and \cite{reve}. 
The computation of $\gamma_{ab}^{(N)}$ for general $N$ is very tedious
and its feasibility is still under study. Therefore one should try to make
an estimate of the three-loop anomalous dimensions. For this estimate one
can use all the information on super-symmetric and conformal relations
available in the literature \cite{bukh}. One also knows the leading and 
sub-leading pole
terms of the type $\gamma_{ab}^{(N)}\sim 1/(N-1)^m$ from the solution of
the BFKL characteristic function \cite{bfkl} which is known up to NLO. 
One also gets
some input from the behaviour at large $N$ which is conjectured for
the ${\overline {\rm MS}}$-scheme in \cite{gly}. According to the last reference
the anomalous dimension behaves like $\gamma_{ab}^{(N)}\sim \ln N$ in all 
orders
of $\alpha_s$ for $N \rightarrow \infty$. Another interesting question is
whether in the ${\overline {\rm MS}}$-scheme the NNLO corrected structure 
functions
are dominated by the coefficient functions rather than the anomalous dimensions
so that the latter play a subordinate role. Using some of the ideas above 
one has already
made a NNLO analysis of the structure function $F_3(x,Q^2)$ in \cite{kps}.
One of the interesting results is that higher twist ($\tau =4$, see Eq. 
(\ref{eqn:6})) contributions get smaller when NNLO corrections are included.
Recently a similar analysis was performed for the singlet and non-singlet
part of the structure function $F_2(x,Q^2)$ in electro-production 
in \cite{sayn}.
\begin{figure}
\begin{center}
  \begin{picture}(100,100)(0,0)
  \DashLine(0,30)(100,30){4}
  \Photon(40,0)(40,30){2}{5}
  \GlueArc(75,20)(55,170,100){5}{12}
  \ArrowLine(65,75)(90,100)
  \ArrowLine(90,50)(65,75)
  \Text(0,25)[t]{$q$}
  \Text(48,5)[t]{$V$}
  \Text(95,100)[t]{$Q$}
  \Text(95,50)[t]{$\bar Q$}
\end{picture}
\end{center}
\begin{center}
  \begin{picture}(100,100)(0,0)
  \DashLine(0,30)(100,30){4}
  \Photon(40,0)(40,30){2}{5}
  \GlueArc(75,20)(55,170,150){5}{4}
  \GlueArc(75,20)(55,119,100){5}{3}
  \ArrowArc(36,60)(15,230,50)
  \ArrowArc(36,60)(15,50,225)
  \Text(0,25)[t]{$q$}
  \Text(48,5)[t]{$V$}
  \Text(51,49)[t]{$Q$}
  \Text(21,83)[t]{$\bar Q$}
\end{picture}
\end{center}
\begin{center}
  \begin{picture}(100,100)(0,0)
  \DashLine(0,30)(100,30){4}
  \Photon(50,0)(50,30){2}{5}
  \GlueArc(50,30)(35,180,115){5}{7}
  \GlueArc(50,30)(35,65,0){5}{7}
  \ArrowArc(50,60)(15,190,10)
  \ArrowArc(50,60)(15,10,190)
  \Text(0,25)[t]{$q$}
  \Text(58,5)[t]{$V$}
  \Text(50,42)[t]{$Q$}
  \Text(50,90)[t]{$\bar Q$}
\end{picture}
\caption[]{\sf Heavy flavour production $V + q \rightarrow q + Q + \bar Q$
 including virtual corrections ($V=\gamma,W,Z$).}
\label{fig5}
\end{center}
\end{figure}
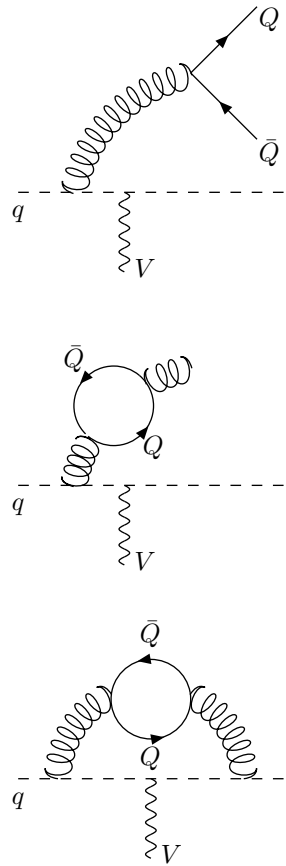

\subsection{Measurement of $\alpha_s$}

It is expected that NNLO corrections to the structure functions diminish
the theoretical error on $\alpha_s$ by about 50 \% with respect to NLO.
There is also a strong dependence on the cut $Q_{cut}^2$ imposed on the data
set \cite{vogt}. A larger cut leads to a decrease of $\alpha_s(M_Z^2)$ by 
about 0.002.
Another issue is the treatment of heavy flavour contributions to the
coupling constant. Usually this is done by imposing the following matching
conditions
\begin{eqnarray}
\label{eqn:27}
\alpha_s(n_f,\Lambda_{n_f},a\,m^2)=\alpha_s(n_f+1,\Lambda_{n_f+1},a\,m^2)\,,
\end{eqnarray}
where $m$ is the mass of the heavy flavour. In the literature one takes $a=1$ 
which means that for $Q^2 > m^2$ the heavy flavour behaves as a massless 
quark. However this does not follow from perturbation theory. An example is 
shown in \cite{blne2}. The heavy flavour contribution to the structure function
appears for the first time in order $\alpha_s^2$ (see Fig. \ref{fig5}) 
provided we assume
that the initial hadron state does not contain a heavy quark. The easiest
way to study these contributions is to look at sum rules e.g. the Gross-
Llewellyn Smith sum rule \cite{grls} which up to order $\alpha_s^2$ 
is 
given by
\begin{eqnarray}
\label{eqn:28}
&&\int_0^1dx\, \Big [ F_3^{\bar \nu p}(x,Q^2)+F_3^{\nu p}(x,Q^2) \Big ]=
\nonumber\\[1ex]
&& 6 \Big [1 + \frac{\alpha_s(n_f,Q^2)}{\pi}\,a_1+
\left (\frac{\alpha_s(n_f,Q^2)}{\pi}\right )^2 \left \{a_2 \right.
\nonumber\\[1ex]
&& \left. +H\left (\frac{Q^2}{m^2} \right) \right \} \Big ]\,.
\end{eqnarray}
Here $a_1$, $a_2$ represent
the light quark and gluon contributions \cite{zn}, \cite{gola}. 
The heavy quark contribution
denoted by $H$ \cite{blne2} behaves asymptotically as
\begin{eqnarray}
\label{eqn:29}
H\left (\frac{Q^2}{m^2} \right )
& {\raisebox{-2 mm}{$\,\stackrel{\rightarrow }
{{\scriptstyle m^2 \gg Q^2 }}\, $} }& \frac{Q^2}{m^2}\,,
\nonumber\\[1ex]
H\left (\frac{Q^2}{m^2} \right )
& {\raisebox{-2 mm}{$\,\stackrel{\rightarrow }
{{\scriptstyle Q^2 \gg m^2 }}\, $} }& \frac{1}{6} a_1 \ln \frac{Q^2}{m^2} + b_2
\,.
\end{eqnarray}
The asymptotic behaviour for $m^2 \gg Q^2$ follows from the decoupling
theorem so that at small scales the heavy quark does not contribute to
the sum rule. The asymptotic behaviour for $Q^2 \gg m^2$ can be interpreted
that the heavy flavour behaves like a light quark which is revealed by
the mass singular logarithm. One can check that the exact expression
becomes equal to the asymptotic one for $Q^2 > 40\,m^2$ which implies
that $a$ in Eq. (\ref{eqn:27}) should be chosen to be 40 instead of 1.
Only for this value it makes sense to re-sum the large logarithm on the 
right-hand side in Eq. (\ref{eqn:29}). This is achieved by absorbing
the logarithm $\ln (Q^2/m^2)$ into the running coupling constant
\begin{eqnarray}
\label{eqn:30}
\alpha_s(n_f+1,Q^2)=\frac{\alpha_s(n_f,Q^2)}{1- \frac{\alpha_s(n_f,Q^2)}{6\pi}
\ln \frac{Q^2}{m^2}}\,.
\end{eqnarray}
After having removed the function $H$ on the righthand side of Eq. 
(\ref{eqn:28}), $\alpha_s(n_f,Q^2)$ has to be replaced by 
$\alpha_s(n_f+1,Q^2)$ so that the perturbation series for the sum rule is 
represented in an $n_f+1$ flavour scheme.
The lesson which one can draw from this calculation is that it is not the
virtuality $p^2$ of the gluon, which is coupled to the heavy quark anti-quark
pair in Fig. \ref{fig5}, but the value of the external kinematic scale $Q^2$
which is relevant for the heavy flavour threshold in the running coupling 
constant and the value of a in Eq. (\ref{eqn:27}).
Of course it might happen that $p^2$ represents a physical observable like the
momentum transfer in e.g. quark-quark scattering. Here it turns out that
if $-p^2 \ge m^2$ the flavour has to be treated as a light quark.
\begin{figure}
\begin{center}
  \begin{picture}(80,80)(0,0)
  \DashArrowLine(0,40)(45,40){4}
  \Photon(45,80)(45,40){2}{5}
  \ArrowLine(45,40)(80,0)
  \Text(0,55)[t]{$s(d)$}
  \Text(80,15)[t]{$c$}
  \Text(55,80)[t]{$W$}
\end{picture}
\caption[]{\sf Charm production $W + s(d) \rightarrow c$.}
\label{fig6}
\end{center}
\end{figure}
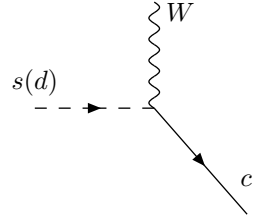

\subsection{Down and strange quark densities at large $x$}
\vspace*{3mm}
One of the goals of future HERA experiments is to provide us with
a better determination of the down $d(x)$ and strange $s(x)$ quark densities.
To start with the former the ratio $d(x)/u(x)$ can be extracted from
the cross sections of the charged current interactions from the ratio
\begin{eqnarray}
\label{eqn:31}
\frac{\sigma_{CC}(e^+~p)}{\sigma_{CC}(e^-~p)}\,.
\end{eqnarray}
It is not excluded that one can expect $d(x)/u(x) \rightarrow {\rm const.}$ 
when $x \rightarrow 1$ which is in contrast with present parametrisations.
Notice that until now information on $d(x)$ is available from the following
observables.
\begin{itemize}
\item[a.]
The first quantity is given by
\begin{eqnarray}
\label{eqn:32}
\frac{F_2^{\mu n}(x,Q^2)}{F_2^{\mu p}(x,Q^2)} \,.
\end{eqnarray}
Until now the structure functions above are extracted from fixed target 
experiments so that
one has to correct for nuclear binding effects \cite{boya}. 
This means that $d(x)$
might deviate from the parametrisations existing in the literature.
\item[b.]
Another observable is the lepton asymmetry in $W$-production in proton 
anti-proton colliders given by the reaction
\begin{eqnarray}
\label{eqn:33}
p + \bar p &\rightarrow& W^+\,(W^-) + `X'
\nonumber\\
&&\mid
\nonumber\\
&&\rightarrow l^+ \nu_l\, (l^- \bar \nu_l) \,.
\end{eqnarray}
\end{itemize}
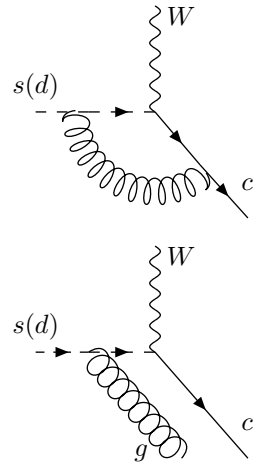
\begin{figure}
\begin{center}
  \begin{picture}(80,80)(0,0)
  \DashLine(0,40)(20,40){4}
  \DashArrowLine(20,40)(45,40){4}
  \Photon(45,80)(45,40){2}{5}
  \GlueArc(45,40)(30,310,179){5}{12}
  \ArrowLine(45,40)(62,20)
  \ArrowLine(62,20)(80,0)
  \Text(0,55)[t]{$s(d)$}
  \Text(80,15)[t]{$c$}
  \Text(55,80)[t]{$W$}

\end{picture}
\end{center}
\begin{center}
  \begin{picture}(80,80)(0,0)
  \DashArrowLine(0,40)(20,40){4}
  \DashArrowLine(20,40)(45,40){4}
  \Photon(45,80)(45,40){2}{5}
  \Gluon(20,40)(55,0){5}{8}
  \ArrowLine(45,40)(80,0)
  \Text(0,55)[t]{$s(d)$}
  \Text(80,15)[t]{$c$}
  \Text(40,5)[t]{$g$}
  \Text(55,80)[t]{$W$}
\end{picture}
\caption[]{\sf Charm production $W + s(d) \rightarrow c + g$
 including radiative corrections.}
\label{fig7}
\end{center}
\end{figure}

The strange quark density can be measured in charm quark production in
the charged current process $e^- (e^+) + p \rightarrow \nu_e (\bar \nu_e) 
+ `X'$. The basic reactions are given by (Fig. \ref{fig6})
\begin{eqnarray}
\label{eqn:34}
s + W^+ \rightarrow c \,, \qquad \bar s + W^- \rightarrow \bar c \,.
\end{eqnarray}
It is found that the radiative corrections to the processes above
are very small. The latter are represented by the one-loop corrections
to reactions (\ref{eqn:34}) and the gluon bremsstrahlung (Fig. \ref{fig7})
\begin{eqnarray}
\label{eqn:35}
s + W^+ \rightarrow c +g \,, \qquad \bar s + W^- \rightarrow \bar c + g \,.
\end{eqnarray}
However the determination of the strange quark density is hampered
by the following backgrounds. The first one is given by the process
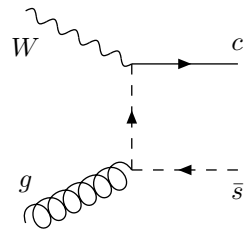
\begin{figure}
\begin{center}
  \begin{picture}(80,80)(0,0)
  \DashArrowLine(80,20)(40,20){4}
  \ArrowLine(40,60)(80,60)
  \DashArrowLine(40,20)(40,60){4}
  \Photon(0,80)(40,60){2}{5}
  \Gluon(0,0)(40,20){5}{6}
  \Text(0,18)[t]{$g$}
  \Text(0,70)[t]{$W$}
  \Text(80,70)[t]{$c$}
  \Text(80,15)[t]{$\bar s$}
\end{picture}
\caption[]{\sf Charm production in $W + g \rightarrow c + \bar s$
              (gluon-boson fusion).}
\label{fig8}
\end{center}
\end{figure}
\begin{eqnarray}
\label{eqn:36}
d + W^+ \rightarrow c \,.
\end{eqnarray}
Although this process is Cabibbo suppressed it is enhanced when it
appears in $e^+ + p \rightarrow \bar \nu_e + `X'$ because in this 
case the d becomes a valence quark and $d_v(x) \gg s(x)$. Another
background is the gluon-boson fusion reaction (Fig. \ref{fig8})
\begin{eqnarray}
\label{eqn:37}
g + W^+ \rightarrow c + \bar s \,, \qquad g + W^- \rightarrow \bar c + s \,,
\end{eqnarray}
which dominates the cross section $d\sigma/dx$ for $x<0.1$. Higher
order corrections to the reaction above \cite{bune} are 
quite appreciable in particular for $F_3(x,Q^2)$ but less for $F_2(x,Q^2)$.
Therefore one can only determine $s(x)$ when the gluon density
is accurately known in the region $5.10^{-3}<x<0.1$. To circumvent
this problem the authors in \cite{krst} have proposed to measure the $D$-mesons
which emerge from the (anti)charm quarks. Instead of $d\sigma/dx$
one has to compute
\begin{eqnarray}
\label{eqn:38}
\frac{d^2 \sigma}{d~x~d~z} \sim  \sum_a f_a \otimes H_a^c \otimes D_c\,,
\end{eqnarray}
where $ H_a^c$ ($a=q,g$) denote the heavy quark coefficient functions
corresponding to reactions (\ref{eqn:34}),(\ref{eqn:35}) and (\ref{eqn:37}).
The function $D_c(z)$ describes the fragmentation of the $c$-quark into the
$D$-meson with
$p_D=z~p_c$. It appears that for $z<0.2$ the gluon-boson fusion dominates
charm quark production so that one has to impose a cut on $z$ in order
to suppress this background. In this case one obtains the quantity
\begin{eqnarray}
\label{eqn:39}
\frac{d \sigma}{d~x} =  \int_{0.2}^1 dz~ \frac{d^2 \sigma}{d~x~d~z} \,,
\end{eqnarray}
leading to the result $d~\sigma^{NLO} \sim d~\sigma^{LO}$ so that the 
Born reaction in Eq. (\ref{eqn:34}) becomes dominant and the extraction
of the strange quark density is possible.

\subsection{Factorization and renormalization scale dependence}
\vspace*{3mm}
As one knows physical quantities like structure functions and cross
sections are scheme independent which implies that they do not depend on
the choice made for the factorization scale $\mu_F$ and the renormalization 
scale $\mu_R$. However finite order perturbation theory violates
this property as one can see as follows. Take as an example the $N$th
moment of a structure function given by
\begin{eqnarray}
\label{eqn:40}
&& F^{(N)}(Q^2)=
\nonumber\\[1ex]
&& f_q^{(N)}\left (\frac{Q_0^2}{\mu^2},a_s(\mu^2)\right )
{\cal C}_q^{(N)}\left (\frac{Q^2}{\mu^2},a_s(\mu^2)\right )\,,
\end{eqnarray}
with the quark density (for the definition of $a_s$ see 
Eq. (\ref{eqn:26}))
\begin{eqnarray}
\label{eqn:41}
f_q^{(N)}&=&\Big [ 1 + a_s(\mu^2) \Big (A_q^{(1)}-c_q^{(1)} \Big ) \Big ] 
\nonumber\\[1ex]
&& \times 
\left [\frac{ a_s(\mu^2)}{a_s(Q_0^2)} \right ]^{\gamma_{qq}^{(0)}/2 \beta_0} 
f_q^{(N)}(Q_0^2)\,.
\end{eqnarray}
and the coefficient function
\begin{eqnarray}
\label{eqn:42}
{\cal C}_q^{(N)}&=&\Big [ 1 + a_s(Q^2)A_q^{(1)} + a_s(\mu^2) \Big (c_q^{(1)}
\nonumber\\[1ex]
&& -A_q^{(1)}\Big ) \Big ]
\left [\frac{ a_s(Q^2)}{a_s(\mu^2)} \right ]^{\gamma_{qq}^{(0)}/2 \beta_0}\,,
\end{eqnarray}
Notice that we have chosen $\mu_F=\mu_R=\mu$ for simplicity. Further
$A_q^{(1)}$ is a scheme independent combination of the the anomalous
dimensions and the coefficient $c_q^{(1)}$. The latter is the order
$\alpha_s$ contribution to the series expansion
of the coefficient function ${\cal C}_q^{(N)}$ and is 
scheme dependent. The lowest order terms in the perturbation series for
the anomalous dimension and the beta-function are given by 
$\gamma_{qq}^{(0)}$ and
$\beta_0$ respectively. When we multiply the coefficient function by
the quark density all scheme dependence cancels up to order $a_s$.
However the order $a_s^2$ term depends on the scheme and the scale $\mu$
because of the missing two-loop coefficient function and the three-loop
anomalous dimension. If one varies the scale like what is usually done
by $Q/2 < \mu < 2~Q$ then the variation in $F^{(N)}$ is wholly due
to the spurious order $a_s^2$ term. Therefore one can raise the question
whether this scale variation is a good estimate of the theoretical error 
due to the absence of higher order QCD corrections. Moreover the scale
variation performed on NLO quantities only probes the effect of the 
Born process on the
NLO corrected cross section. The latter has the typical form
\begin{eqnarray}
\label{eqn:43}
\hat \sigma_{ab}^{(1)}&=& \bar \sigma_{ab}^{(1)} + \frac{1}{2}P_{cb}^{(0)}
\otimes \hat \sigma_{ac}^{(0)} \ln \left( \frac{Q^2}{\mu_F^2} \right )
\nonumber\\[1ex]
&& -\beta_0 \hat \sigma_{ab}^{(0)} \ln \left( \frac{Q^2}{\mu_R^2} \right )\,.
\end{eqnarray}
Here $\hat \sigma_{ab}^{(0)}$ and $\hat \sigma_{ab}^{(1)}$ represent
the lowest (Born) and the first order cross section respectively. The
lowest order splitting function is denoted by $P_{cb}^{(0)}$.
The quantity $\bar \sigma_{ab}^{(1)}$ contains all first order contributions
originating from new production mechanism's. From Eq. (\ref{eqn:43}) one 
infers that all scale dependent logarithms are multiplied by the 
Born cross section only.
In the case that $\bar \sigma_{ab}^{(1)}$ is large the latter gives
a better indication about the theoretical error due to higher order
corrections than a simple scale variation of the NLO hadronic cross section.

\section{Re-summation of large corrections occurring at small and large $x$}
\vspace*{3mm}
From the computations of splitting functions $P_{ab}(x)$ and
coefficient functions ${\cal C}_{k,a}(x,Q^2/\mu^2)$ we infer that
large corrections can appear at 
\begin{itemize}
\item[a.] $x \rightarrow 0$ ~~~~(soft gluon exchanges)
\item[b.] $x \rightarrow 1$ ~~~~(soft gluon bremsstrahlung)
\end{itemize}
However the relevance of these corrections does not only depend on
the quantities above but also on the behaviour of the parton densities.
This one can see by a study of the structure functions
given by
\begin{eqnarray}
\label{eqn:44}
&& F_k(x,Q^2) \sim 
\nonumber\\[1ex]
&& \sum_a \int_x^1 \frac{dz}{z}~f_a \left (\frac{x}{z},\mu^2 \right )
{\cal C}_{k,a}\left (z,\frac{Q^2}{\mu^2} \right )\,.
\end{eqnarray}
Because of the convolution integral it might happen that $f_a(x/z,\mu^2)$
vanishes in the regions $z \rightarrow 0$ or $z \rightarrow 1$ so that
the total contribution to the structure functions is small in spite of the fact
that the coefficient functions and the splitting functions blow up
in these regions. Therefore a re-summation of these non-dominant corrections
will lead to an overestimate of their effect on the physical quantities.
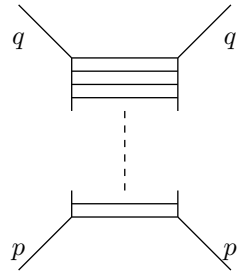
\begin{figure}
\begin{center}
  \begin{picture}(80,100)(0,0)
  \Line(0,100)(20,80)
  \Line(60,80)(80,100)
  \Line(20,80)(60,80)
  \Line(20,80)(20,60)
  \Line(60,80)(60,60)

  \Line(20,75)(60,75)
  \Line(20,70)(60,70)
  \Line(20,65)(60,65)

  \DashLine(40,60)(40,30){3}
  \Line(20,20)(20,30)
  \Line(60,20)(60,30)
  \Line(20,25)(60,25)
  \Line(60,20)(80,0)
  \Line(20,20)(60,20)
  \Line(0,0)(20,20)

  \Text(0,90)[t]{$q$}
  \Text(80,90)[t]{$q$}
  \Text(0,10)[t]{$p$}
  \Text(80,10)[t]{$p$}
\end{picture}
\caption[]{\sf Ladder graph in $\Phi_6^3$-theory.}
\label{fig9}
\end{center}
\end{figure}
\subsection{Large corrections at small $x$}
\vspace*{3mm}
The large corrections occurring for $z \rightarrow 0$ are due to soft
gluon exchanges in the $t$-channel. They have the following form 
\begin{eqnarray}
\label{eqn:45}
&& {\cal C}_{k,a}^{(l)} (z,1) \sim  \frac{\ln^{l-2}z}{z} \qquad l \ge 2 \,,
\nonumber\\[1ex]
&& P_{ab}^{(l)}(z) \sim  \frac{\ln^l z}{z} \qquad l \ge 0\,.
\end{eqnarray}
Because the singular behaviour is due to gluon exchange it only
appears in the singlet part of the coefficient functions and splitting
functions.
The re-summation of these logarithms is performed by the BFKL equation
\cite{bfkl}.
However one has to be careful with the interpretation of the solution
of this equation which will be defined as ${\cal C}^{BFKL}(z)$. One
cannot claim that the latter represents the asymptotic form of the
exact coefficient function ${\cal C}^{EXACT}(z)$ in the limit 
$z \rightarrow 0$. This can be shown in $\Phi_6^3$-theory where the exact
solution  \cite{lov} is known for the ladder diagrams (see Fig. \ref{fig9}). 
In this case one can also re-sum the leading terms given at each order 
\cite{blne1} which can be considered as the analogue of the BFKL-solution 
but now in the case of $\Phi_6^3$-theory.
However this re-summation
leads to an answer which differs by a factor of two and larger from the exact
result taken in the limit $z \rightarrow 0$. This means that also 
non-leading terms determine the asymptotic expression for the ladder
graphs. Since the non-leading order terms also appear in non-planar
diagrams, which do not belong to the class of ladder graphs, one cannot even 
claim that the asymptotic expression for the total coefficient function
is determined by the ladder approximation only. Since the exact
expression for the re-summation of a subset of graphs, analogous to the
ladder approximation, is unknown in QCD it is very hard to judge what the 
relation is between the BFKL solution and the perturbative QCD result in the small 
$x$-region.

\subsection{Large corrections at large $x$}
\vspace*{3mm}
The large corrections which appear in the limit $z \rightarrow 1$
are due to soft gluon radiation. They have the following form when the 
$l$th order quantities are presented in the ${\overline {\rm MS}}$-scheme for
$\mu^2=Q^2$
\begin{eqnarray}
\label{eqn:46}
&& {\cal C}_{k,a}^{(l)} (z,1) \sim  \left (\frac{\ln^{2l-1}(1-z)}{1-z}\right )_+ 
\qquad l \ge 1 \,,
\nonumber\\[1ex]
&& P_{ab}^{(l)}(z) \sim  \left (\frac{1}{1-z} \right )_+ 
\qquad l \ge 0\,.
\end{eqnarray}
Notice that the expression for the splitting function is a conjecture
made in \cite{gly}. The singularity structure given in the equation above is
typical for the non-singlet coefficient function contributing to
$F_2$ and $F_3$. If the coefficient function in the above equation is 
convoluted with regular functions the singularity structure becomes much 
milder. In this case they have the form
\begin{eqnarray}
\label{eqn:47}
 {\cal C}_{k,a}^{(l)} (z,1) \sim a_l \ln^{2l-2} (1-z) \qquad l \ge 2\,.
\end{eqnarray} 
Examples of this behaviour can be found for the coefficient functions
contributing to the longitudinal structure function $F_L$ and the
cross section for direct photon production. The study of the 
logarithms which appear in the coefficient function of Eq. (\ref{eqn:47})
has a twofold purpose.
\begin{figure}
\begin{center}
  \begin{picture}(80,90)(0,0)
  \Gluon(14,15)(65,15){4}{6}
  \Photon(40,90)(40,70){2}{4}
  \ArrowLine(10,0)(40,70)
  \ArrowLine(40,70)(70,0)
  \Text(47,90)[t]{$q$}
  \Text(5,0)[t]{$p_1$}
  \Text(75,0)[t]{$p_2$}
\end{picture}
\caption[]{\sf One-loop vertex graph contributing to
        $F^{(1)}(Q^2)$.}
\label{fig10}
\end{center}
\end{figure}
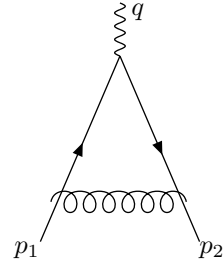

\begin{itemize}
\item[a.] 
Prediction of the coefficients $a_l$ in Eq. (\ref{eqn:47})
and a comparison with the results obtained from the exact expression.
In this way one can check the exact calculation.
\item[b.] 
The investigation of the coefficients $a_l$ will reveal how
one has to re-sum the large logarithms.
\end{itemize}
First we would like to discuss the work done in \cite{ass} to re-sum the 
logarithmic
terms occurring in the longitudinal coefficient function contributing
to $F_L$. Up to $l=2$ one obtains the following form when the 
scales are chosen to be $\mu_F=\mu_R=Q$
\begin{eqnarray}
\label{eqn:48}
&& {\cal C}_{L,q}^{(0)} (z,1)=0\,, \qquad {\cal C}_{L,q}^{(1)} (z,1)=C_F~z\,,
\nonumber\\[1ex]
&& {\cal C}_{L,q}^{(2)} (z,1)= C_F^2 \Big [ \frac{1}{2}\ln^2 (1-z)
+ (\frac{9}{4}-2\zeta(2)) 
\nonumber\\[1ex]
&& \times \ln (1-z) \Big ]+ C_AC_F \Big [ (\zeta(2)-1)
\ln (1-z)\Big ] 
\nonumber\\[1ex]
&& - \frac{1}{4} \beta_0 C_F \ln (1- z) + \mbox{regular terms in $z$}\,,
\end{eqnarray}
Here $C_F$ and $C_A$ represent the well-known colour factors. The re-summation
of these logarithms is easier when one performs the Mellin transformation.
In the limit $N \rightarrow \infty$ one obtains
\begin{eqnarray}
\label{eqn:49}
&& {\cal C}_{L,q}^{(N),(2)} (z,1)= \frac{C_F}{2N} \Big [ \gamma_K^{(0)}
\ln^2 \frac{N}{N_0} 
\nonumber\\[1ex]
&&  - (\gamma_{J'}^{(1)} - \frac{1}{2} \beta_0) \ln \frac{N}{N_0} \Big ]\,,
\end{eqnarray}
and $\gamma_{J'}$ is the anomalous dimension of a certain operator
determining the jet function \cite{ass}. The quantity $\gamma_K$ represents
the Sudakov anomalous dimension and it is related to the infrared structure
of a gauge theory like QCD. It also occurs in the vertex correction to
the quark-gluon vertex represented by $F(Q^2)$. Using $n$-dimensional 
regularization with $\varepsilon =n-4$ and $p_1^2=p_2^2=0$ 
one obtains for the one-loop correction (see Fig. \ref{fig10})
\begin{eqnarray}
\label{eqn:50}
F^{(1)}(Q^2)= \left (\frac{Q^2}{\mu^2} \right )^{\varepsilon/2}
\Big [ \frac{\gamma_K^{(0)}}{\varepsilon^2} + \cdots \Big ] \,.
\end{eqnarray}
The two-loop expression (Fig. \ref{fig11}) looks like
\begin{eqnarray}
\label{eqn:51}
&& F^{(2)}(Q^2)= \left (\frac{Q^2}{\mu^2} \right )^{\varepsilon}
\Big [ \frac{(\gamma_K^{(0)})^2}{2\varepsilon^4} + 
(\quad) \frac{1}{\varepsilon^3}
\nonumber\\[1ex]
&& + \frac{\gamma_K^{(1)}+\cdots }{2\varepsilon^2}
+ \cdots \Big ] \,.
\end{eqnarray}
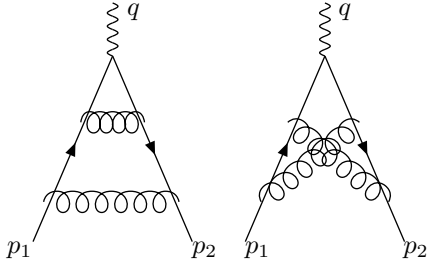
\begin{figure}
\begin{center}
  \begin{picture}(160,90)(0,0)
  \Gluon(28,45)(52,45){4}{4}
  \Gluon(14,15)(65,15){4}{6}
  \Photon(40,90)(40,70){2}{4}
  \ArrowLine(10,0)(40,70)
  \ArrowLine(40,70)(70,0)
  \Gluon(106,45)(144,15){4}{6}
  \Gluon(96,15)(133,45){4}{6}
  \Photon(120,90)(120,70){2}{4}
  \ArrowLine(90,0)(120,70)
  \ArrowLine(120,70)(150,0)
  \Text(128,90)[t]{$q$}
  \Text(48,90)[t]{$q$}
  \Text(5,0)[t]{$p_1$}
  \Text(75,0)[t]{$p_2$}
  \Text(95,0)[t]{$p_1$}
  \Text(155,0)[t]{$p_2$}
\end{picture}
\caption[]{\sf Some two-loop vertex graphs contributing to
        $F^{(2)}(Q^2)$.}
\label{fig11}
\end{center}
\end{figure}
From the coefficients of the leading pole terms one infers that
the re-summed Sudakov vertex correction gets an exponential form \cite{sen}, 
\cite{col}.
Because of the analogy between the $\ln N$ terms in the coefficient
function and the $1/\varepsilon$ poles in $F(Q^2)$ we expect that
the same will happen in the former case. The re-summed expression 
for Eq. (\ref{eqn:49}) has been found in \cite{ass}.

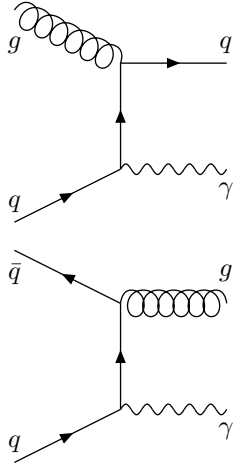
\begin{figure}
\begin{center}
  \begin{picture}(80,80)(0,0)
  \ArrowLine(40,60)(80,60)
  \ArrowLine(40,20)(40,60)
  \ArrowLine(0,0)(40,20)
  \Gluon(0,80)(40,60){5}{6}
  \Photon(40,20)(80,20){2}{5}
  \Text(0,10)[t]{$q$}
  \Text(0,70)[t]{$g$}
  \Text(80,70)[t]{$q$}
  \Text(80,15)[t]{$\gamma$}
\end{picture}
\end{center}
\begin{center}
  \begin{picture}(80,80)(0,0)
  \ArrowLine(40,60)(0,80)
  \ArrowLine(40,20)(40,60)
  \ArrowLine(0,0)(40,20)
  \Photon(40,20)(80,20){2}{5}
  \Gluon(40,60)(80,60){5}{6}
  \Text(0,10)[t]{$q$}
  \Text(0,75)[t]{$\bar q$}
  \Text(80,75)[t]{$g$}
  \Text(80,15)[t]{$\gamma$}
\end{picture}
\caption[]{\sf The Born contributions to direct photon production
        $q + g \rightarrow q + \gamma$, $q + \bar q \rightarrow g + \gamma$.}
\label{fig12}
\end{center}
\end{figure}

A similar attempt to re-sum the large logarithmic terms has been made
for the partonic cross sections (coefficient functions) which appear in 
direct photon production given by
\begin{eqnarray}
\label{eqn:52}
&& H_1(p_1)+H_2(p_2) \rightarrow \gamma(E_T) + `X' \,,
\nonumber\\[1ex]
&& S=(p_1+p_2)^2\,, \qquad x=\frac{2~E_T}{\sqrt S}\,.
\end{eqnarray}
In the reaction above $H_1$ and $H_2$ denote the incoming hadrons and $E_T$ 
is the transverse energy of the photon.  
For this process the $l$th order finite partonic cross section has the form
\begin{eqnarray}
\label{eqn:53}
&& \hat \sigma_{ab}^{(l)}(z) \sim \hat \sigma_{ab}^{(0)}(z) \Big [ d_{l,2l} 
\ln^{2l} (1-z)  
\nonumber\\[1ex]
&&+  d_{l,2l-1}\ln^{2l-1} (1-z) + \cdots \Big ] \qquad l \ge 1 \,,
\end{eqnarray}
where $\sigma_{ab}^{(0)}$ denote the Born cross sections corresponding
to the processes in Fig. \ref{fig12}. 
Notice that $\sigma_{qg}^{(0)}$ is dominant
in $p-N$ scattering whereas $\sigma_{q \bar q}^{(0)}$ becomes more important
in $p-\bar p$ collisions. Like in the case of the longitudinal
coefficient function it is more convenient
to take the Mellin transform of the partonic cross sections which become
\begin{eqnarray}
\label{eqn:54}
&& \hat \sigma_{ab}^{(N),(l)} \sim \hat \sigma_{ab}^{(N),(0)} \Big [ c_{l,2l}
\ln^{2l} \frac{N}{N_0}
\nonumber\\[1ex]
&&+  c_{l,2l-1}\ln^{2l-1}\frac{N}{N_0} + \cdots \Big ]\,, \quad l \ge 1\,.
\end{eqnarray}
The re-summation of the $\ln N$ terms is carried out in \cite{cat}
(see also \cite{los}, \cite{kid})
and leads to exponentiation analogous to what we have discussed above.
The goal of this re-summation was to get a better description of the 
transverse energy $E_T$ distribution of the direct photon. It turns out
\cite{cat} that the re-summation leads to an improvement of the 
factorization scale
dependence with respect to the exact NLO hadronic cross section. 
However the discrepancy between $d \sigma^{NLO}/dE_T$
and the data, in particular those coming from the E706 experiment, remains.
This occurs in the region $x <0.57$ where $E_T< 9~{\rm GeV}$ and 
$\sqrt S=31.6 ~{\rm GeV}$. There is also a little discrepancy for the
UA6 experiment in the region $x <0.5$ where $E_T< 6~{\rm GeV}$ and
$\sqrt S=24.3 ~{\rm GeV}$. Notice that these regions lie outside the
large $x$-region which is close to 1 where soft gluons dominate. Therefore one 
cannot expect that soft gluon re-summation will cure the discrepancy found
at smaller $x$. Several proposals have been made to describe the data
at smaller transverse energy. One is made in \cite{mrst} where it is assumed
that the initial state partons have an intrinsic transverse momentum. Another
point of view is presented in \cite{aur} claiming that the low $E_T$ data
are inconsistent among the various experiments.
\section{Conclusions}
\vspace*{3mm}
Summarizing this talk we conclude
\begin{itemize}
\item[1.]
Since HERA started to explore the small $x$-region one has seen a
resurgence of models which were ruled out by fixed target deep inelastic
experiments carried out at large $x$ i.e. $0.01<x<1$. Some of
them are discussed like for instance the Regge Pole Model and Vector
Meson Dominance. The BFKL approach which became popular with the advent
of HERA data merges ideas taken from the Regge Pole Model as well as from QCD.
None of these models is able to predict the $Q^2$-evolution of the deep
inelastic structure functions contrary to QCD where the evolution follows
from the renormalization group equations. However until now the
$x$-dependence of the structure function cannot be predicted by QCD.
Therefore as far as the small $x$-behaviour is concerned the aforementioned
models cannot be ruled out since they give an equally good description of
the data.
\item[2.]
Future experiments at HERA and fixed target experiments at FNAL will
explore the large $x$-region ($x>0.01$) where there is no alternative model
to perturbative QCD. 
\item[3.]
The future deep inelastic lepton-hadron experiments have to lead to 
precision tests of QCD akin to the program carried out for $e^+~e^-$
colliders like LEP and SLC. In particular one has to perform,
\begin{itemize}
\item[a.]
An accurate determination of the strong coupling constant $\alpha_s$
which can be obtained from structure functions as well as jet distributions.
\item[b.]
A study of heavy flavour thresholds in $\alpha_s$ and the structure
functions $F_k(x,Q^2)$.
\item[c.]
A better estimate of large corrections in the perturbation series for 
physical quantities.
\item[d.]
The computation of the three-loop splitting functions (anomalous dimensions) 
in order to give a full NNLO analysis of $F_k(x,Q^2)$.
\end{itemize}
\end{itemize}


\begin{thebibliography}{9}
\bibitem{bloom} E.D. Bloom et al., Phys. Rev. Lett. {\bf 23} (1969) 930;\\
                M. Breidenbach et al., Phys. Rev. Lett. {\bf 23} (1969) 935.
\bibitem{bartel} W. Bartel et al., Phys. Lett. {\bf B28} (1968) 148;\\ 
                 W. Albrecht et al., DESY Report 69/7 (1969).
\bibitem{brpr} R.A. Brandt, G. Preparata, Nucl. Phys. {\bf B27} (1971) 541;\\
               Y. Frishman,  Ann. Phys. (NY) {\bf 66} (1971) 373.
\bibitem{feyn} R.P. Feynman, Phys. Rev. Lett. {\bf 23} (1969) 1415;\\
               S.D. Drell, T.M. Yan, Ann. Phys. (NY) {\bf 66} (1971) 578.
\bibitem{saku} J.J. Sakurai, Phys. Rev. Lett. {\bf 22} (1969) 981;
               C.F. Cho and J.J. Sakurai, Phys. Lett. {\bf B31} (1970) 22.
\bibitem{har} H. Harari, Phys. Rev. Lett. {\bf 22} (1969) 1078;\\
              H.D.I. Abarbanel, M.I. Goldberger, S.B. Treiman,
               Phys. Rev. Lett. {\bf 22} (1969) 500.
\bibitem{bfkl}  E.A. Kuraev, L.N. Lipatov, V.S. Fadin, Sov. Phys. JETP {\bf 45}
                (1977) 199;\\
                Y. Balitskii, L.N. Lipatov, Sov. J. Nucl. Phys. {\bf 28} (1978) 822.
\bibitem{iof} V.N. Gribov, B.L. Ioffe, I.Ya. Pomeranchuk, Sov. J.
              Nucl. Phys. {\bf 2 }(1966) 549;\\
              B.L. Ioffe, Phys. Lett. {\bf B30} (1969) 123.
\bibitem{cagr}  C.G. Callan, D.J. Gross, Phys. Rev. Lett. {\bf 22} (1969) 156.
\bibitem{nieh} H.T. Nieh, Phys. Rev. D1 (1970) 3161, ibid. {\bf D7} (1973) 4301.
\bibitem{sasc} J.J. Sakurai, D. Schildknecht, Phys. Lett. {\bf B40} (1972) 121;
               B. Gorczyca, D. Schildknecht,  Phys. Lett.{\bf B47} (1973) 71.
\bibitem{scsp} D. Schildknecht, H. Spiesberger, hep-ph/9707447.
\bibitem{adloff} C. Adloff et al. (H1-collaboration), Phys. Lett. {\bf B393}
                 (1997) 452.
\bibitem{cudo} J.H. Cudell, A. Donachie, P.V. Landshoff, Phys. Lett. {\bf B448}
               (1999) 281; P.V. Landshoff, these proceedings, 
               hep-ph/9905230. 
\bibitem{blne1} J. Bl\"umlein, W.L. van Neerven, Phys. Lett. {\bf B450} (1999) 412.
\bibitem{blvo} J. Bl\"umlein, A. Vogt, Phys. Rev. {\bf D57} (1998) R1,
               Phys. Rev. {\bf D58} (1998) 014020.
\bibitem{zn} E.B. Zijlstra, W.L. van Neerven, 
             Phys. Lett. B272 (1991) 127, Phys. Lett. {\bf B273} (1991) 476,
             Phys. Lett. B297 (1992) 377, Nucl. Phys. {\bf B383} (1992) 525.
\bibitem{lvr} S.A. Larin, T. van Ritbergen and J.A.M. Vermaseren,
              Nucl. Phys. {\bf B427} (1994) 41;\\
             S.A. Larin et al., Nucl. Phys. {\bf B492} (1997) 338.
\bibitem{begr} J.A. Gracey, Phys. Lett {\bf B322} (1994) 141;\\
               J.A. Benett and J.A. Gracey, Phys. Lett. {\bf B432} (1998) 209,
               Nucl. Phys. {\bf B517} (1998) 241.
\bibitem{blku} J. Blumlein, S. Kurth, hep-ph/9708388, hep-ph/9810241,
               Phys. Rev. {\bf D} in print.
\bibitem{reve} E. Remiddi, J.A.M. Vermaseren,\\ hep-ph/9905237.
\bibitem{gly} A. Gonzalez-Arroyo, C. Lopez, F.J. Yndurain, Nucl. Phys.
              {\bf B153} (1979) 161.
\bibitem{kps} A.L. Kataev, G. Parente and A.V. Sidorov,
             hep-ph/9809500, hep-ph/9904332.
\bibitem{sayn} J. Santiago, F.J. Yndurain, hep-ph/9904344.
\bibitem{vogt} A. Vogt, these proceedings.
\bibitem{bukh}  A.P. Bukhvostov et al., Nucl. Phys. {\bf B258} (1985) 601.
\bibitem{blne2} J. Bl\"umlein, W.L. van Neerven, Phys. Lett. {\bf B450} (1999) 417.
\bibitem{grls} D.J. Gross, C.H. Llewellyn Smith, Nucl. Phys. {\bf B14}(1969) 337.
\bibitem{gola} S.G. Gorishni, S.A. Larin, Nucl. Phys. {\bf B283} (1987) 452.
\bibitem{boya} A. Bodek, U.K. Yang, hep-ph/9809480.
\bibitem{bune} M. Buza, W.L. van Neerven, Nucl. Phys. {\bf B500} (1997) 301.
\bibitem{krst} S. Kretzer and M. Stratmann, hep-ph/9902426.
\bibitem{lov} C. Lovelace, Phys. Lett. {\bf B55} (1975) 187; Nucl. Phys. {\bf B95}
              (1975) 12.
\bibitem{ass} R. Akhoury, G. Sotiropoulos, G. Sterman, hep-ph/9903442.
\bibitem{sen} A. Sen, Phys. Rev. {\bf D27} (1983) 2997.
\bibitem{col} J. C. Collins in "Perturbative Quantum Chromodynamics", Ed.
              A.H. Mueller, pg.573.
\bibitem{cat} S. Catani et al., hep-ph/9903436
\bibitem{los} E. Laenen, G. Oderda, G. Sterman, Phys. Lett. {\bf B438} (1998) 173.
\bibitem{kid} N. Kidonakis, hep-ph/9902484.
\bibitem{mrst} A.D. Martin et al., Eur. Phys. J. {\bf C4} (1998) 463.
\bibitem{aur} P. Aurenche et al., hep-ph/9811382.
\end{thebibliography}
\end{document}